\documentclass[twocolumn]{aastex61}

\usepackage{xspace}

\renewcommand{\rho}{\varrho}

\def \gray {$\gamma$-ray\xspace}
\def \grays {$\gamma$-rays\xspace}
\def \flx {photons $\mathrm{cm}^{-2}$ $\mathrm{s}^{-1}$\xspace}
\def \eflx {erg $\mathrm{cm}^{-2}$ $\mathrm{s}^{-1}$\xspace}
\def \grid {AGILE-\textit{GRID}\xspace}
\def \bat {\textit{Swift}/BAT\xspace}
\def \lat {\textit{Fermi}-LAT\xspace}
\def \fermi {\textit{Fermi}\xspace}
\def \vcyg {V404 Cyg\xspace}

\def \mod {\textbf}

\begin{document}

\title{High-energy gamma-ray activity from V404 Cygni detected by AGILE during the 2015 June outburst}

\correspondingauthor{Giovanni Piano}
\email{giovanni.piano@iaps.inaf.it}

\author{G. Piano}
\affiliation{INAF-IAPS, Via del Fosso del Cavaliere 100, I-00133, Roma, Italy}

\author{P. Munar-Adrover}
\affiliation{INAF-IAPS, Via del Fosso del Cavaliere 100, I-00133, Roma, Italy}

\author{F. Verrecchia}
\affiliation{ASI Data Center (ASDC), Via del Politecnico snc, I-00133, Roma, Italy}
\affiliation{INAF-OAR,  Via di Frascati 33, I-00040 Monte Porzio Catone (RM), Italy}

\author{M. Tavani}
\affiliation{INAF-IAPS, Via del Fosso del Cavaliere 100, I-00133, Roma, Italy}
\affiliation{INFN Roma Tor Vergata, Via della Ricerca Scientifica 1, I-00133, Roma, Italy}
\affiliation{Dipertimento di Fisica, Universit\`{a} di Roma ``Tor Vergata'', Via Orazio Raimondo 18, I-00173 Roma, Italy}

\author{S. A. Trushkin}
\affiliation{Special Astrophysical Observatory RAS, Karachaevo-Cherkassian Republic, Nizhnij Arkhyz 36916, Russian Federation}
\affiliation{Kazan Federal University, Kazan 420008, Russian Federation}

\begin{abstract}

The AGILE satellite detected transient \gray activity from the X-ray binary V404 Cygni, during the June 2015 outburst observed in radio, optical, X-ray and \gray frequencies. The high-energy \gray emission was observed by AGILE in the 50--400 MeV energy band, between 2015-06-24 UT 06:00:00 and 2015-06-26 UT 06:00:00 (MJD 57197.25--57199.25), with a detection significance of ${\sim}4.3{\sigma}$.
The \gray detection, consistent with a contemporaneous observation by \lat, is correlated with a bright flare observed at radio and hard X-ray frequencies, and with a strong enhancement of the 511 keV line emission, possibly indicating plasmoid ejections in a lepton-dominated transient jet. The AGILE observations of this binary system are compatible with a microquasar scenario in which transient jets are responsible for the high-energy \gray emission.

\end{abstract}


\section{Introduction} \label{sec:intro}
V404 Cygni (hereafter \vcyg), also known as GS 2023+338, is a low mass X-ray binary (LMXB) located at a distance of $2.39 \pm 0.14$ kpc, accurately inferred by a parallax measurement \citep{miller-jones_2009}. The system is composed of a $9 ^{+0.2}_{-0.6}$ $M_\odot$ black hole (BH) and a $0.7^{+0.3}_{-0.2}$ $M_\odot$ K3 III companion star with an orbital period of $6.4714 \pm 0.0001$ days \citep{casares_1992} and a $67^{\circ}$ inclination with respect to the line of sight \citep{casares_1994,shahbaz_1994,khargharia_2010}. \vcyg was discovered as a nova during an optical outburst in 1938, and it was observed for the first time in the X-ray band by the GINGA satellite, during an intense outburst in 1989 \citep{makino_1989}.

LMXBs are usually transient systems, showing long periods of quiescence (years), with faint and rapidly variable emission in the X-ray and radio frequencies \citep{miller-jones_2008,hynes_2009,rana_2016}, and bright outburst states (weeks/months). The X-ray luminosities span from $10^{31-33}$ $\mathrm{erg~s^{-1}}$ during quiescence phase up to Eddington limit ($L_{\mathrm{Edd}} \approx 10^{39}$ $\mathrm{erg~s^{-1}}$ for a 9 $M_{\odot}$ BH) in the outburst states \citep{rodriguez_2015}.

After a quiescence period of $\sim$26 years, the detection by \bat triggered the observations of a new active phase on 2015 June 15 (MJD 57188, \citealt{barthelmy_2015}) that lasted $\sim$2 weeks and was observed across all wavelengths (from radio to soft \grays), with a highly variable emission \citep{trushkin_2015,rodriguez_2015,jenke_2016}. The correlated variability between optical and X-ray emissions during the bright June 2015 activity has been interpreted as a consequence of the instability of its large accretion disk (the outer accretion disk radius is $R_{out}\sim 10^7$ km). A disruption of the accreting mass inflow into the inner part of the disk can represent a critical factor to explain the observed scenario of this intense outburst: large-amplitude fluctuations which rapidly ended after only two weeks. The outburst possibly started in the innermost part of the disk, but it was not sustained by the accretion. The mass inflow could be disrupted either by a low surface density in the outer part of the disk -- because of its long orbital period \citep{kimura_2016} -- or by a strong outer-disk wind regulating the accretion \citep{munoz-darias_2016}.

In June 2015, besides optical and X-ray intense variable emissions, indicating the combined activity of the corona-disk system, observations in radio \citep{trushkin_2015} and \gray energies \citep{siegert_2016, loh_2016} confirmed the ejection of relativistic plasma jets. In particular, hints of $e^+ ~ e^-$ pair annihilation, which are consistent with a microquasar scenario, have been found by INTEGRAL during the active phase \citep{siegert_2016}.

In this paper we present the AGILE observations of \vcyg during the peak phase of this intense activity, compare the results with the \lat data, and analyze the high-energy (HE) \gray emission in a multi-wavelength context.

\section{Observations and Data Analysis}\label{sec:agile}

We analyzed the data collected by the \textit{GRID} (Gamma-Ray Imaging Detector, \citealp{barbiellini_2002,prest_2003}), the \gray silicon-tracker imager on board the AGILE satellite (for a detailed description of the AGILE payload: \citealp{tavani_2009a}), and we compared our results with the \lat observations of the system (see Appendix~\ref{appdx:fermi}).

The \grid is sensitive to \gray photons in the energy range 30 MeV -- 30 GeV. The point spread function (PSF) at 100 MeV and 400 MeV is $4.2^{\circ}$ and $1.2^{\circ}$ ($68\%$ containment radius), respectively \citep{sabatini_2015}.
AGILE had operated in a ``pointing'' mode data-taking, characterized by fixed attitude observations, until November 2009, when the satellite entered in ``spinning'' mode, covering a large fraction of the sky with a controlled rotation of the pointing axis. In this current observing mode, typical 2-day integration-time sensitivity (3$\sigma$) for sources in the Galactic plane and photon energy above 100 MeV is ${\sim}10^{-6}$ \flx.

The analysis of the \grid data was carried out with the new \verb+Build_23+ scientific software, \verb+FM3.119+ calibrated filter and \verb+I0025+ response matrices. The consolidated archive, available from the ASI Data Center (\verb+ASDCSTDk+), was analyzed by applying South Atlantic Anomaly event cuts and $80^{\circ}$ Earth albedo filtering. Only incoming \gray events with an off-axis angle lower than $60^{\circ}$ were selected for the analysis. Statistical significance and flux determination of the point sources were calculated by using the AGILE multi-source likelihood analysis (MSLA) software \citep{bulgarelli_2012} based on the Test Statistic (TS) method as formulated by \citealp{mattox_1996}. This statistical approach provides a detection significance assessment of a \gray source by comparing maximum-likelihood values of the null hypothesis (no source in the model) with the alternative hypothesis (point source in the field model). In this work we report $68\%$ confidence level (C. L.) flux upper limits (ULs) if TS $< 9$ (detection significance $\lesssim 3$) and flux values with the corresponding 1$\sigma$ statistical errors otherwise (TS $\geqslant 9$).

Before analyzing the outburst interval, we carried out an analysis of the Cygnus field during a 5-month period between 2015-01-01 UT 12:00:00 and 2015-06-01 UT 12:00:00. The analysis took into account two different energy bands: 50--400 MeV and above 400 MeV. This preliminary analysis allowed us to model the \gray field just before the onset of the strong activity from \vcyg. For both energy ranges we performed a MSLA including -- in addition to \vcyg\  -- the 3 main pulsars of the Cygnus region (PSR J2021+3651, PSR J2021+4026 and PSR J2032+4127), which are known to be intense and persistent \gray sources. We modeled the \gray spectrum for \vcyg by assuming a simple power law with a 2.1 photon index\footnote{This is a standard value used in the AGILE analysis for unknown-spectrum or low-statistics sources (see \citealp{pittori_2009}).}. Flux ULs of $1 \times 10^{-7}$ \flx and $2 \times 10^{-8}$ \flx were found for \vcyg in 50--400 MeV and $>$400 MeV energy bands, respectively. The fluxes of the 3 \gray pulsars, found in these 5-month preliminary analyses, were kept fixed in the following MSLAs for the outburst phase. Furthermore, the diffuse emission (Galactic and extragalactic) quantified in this preliminary analysis was kept fixed during the study of the active period (June 2015).

For the outburst activity period we analyzed the time interval from 2015-06-20 UT 06:00:00 to 2015-06-30 UT 06:00:00, across the giant flare recorded by \bat \citep{segreto_2015} and RATAN-600 \citep{trushkin_2015} on 2015 June 26 (MJD 57199). A 48-hours bin light-curve for \vcyg was calculated (with a MSLA approach) in both energy bands. We selected these period (five 48-hours time intervals) in order to ensure a stable exposure for the target source. In the 50--400 MeV energy band, the on-source exposure for each bin was found to be almost constant around a value of ${\sim}5.16 \times 10^6$ $\mathrm{cm^2~s}$ with a mean fluctuation of ${\sim}3 \%$.
No detection with TS $> 9$ was found in the $>$400 MeV energy band, with 48h flux ULs lower than $5 \times 10^{-7}$ \flx.

\begin{figure}
  \begin{center}
	\includegraphics[width=\columnwidth]{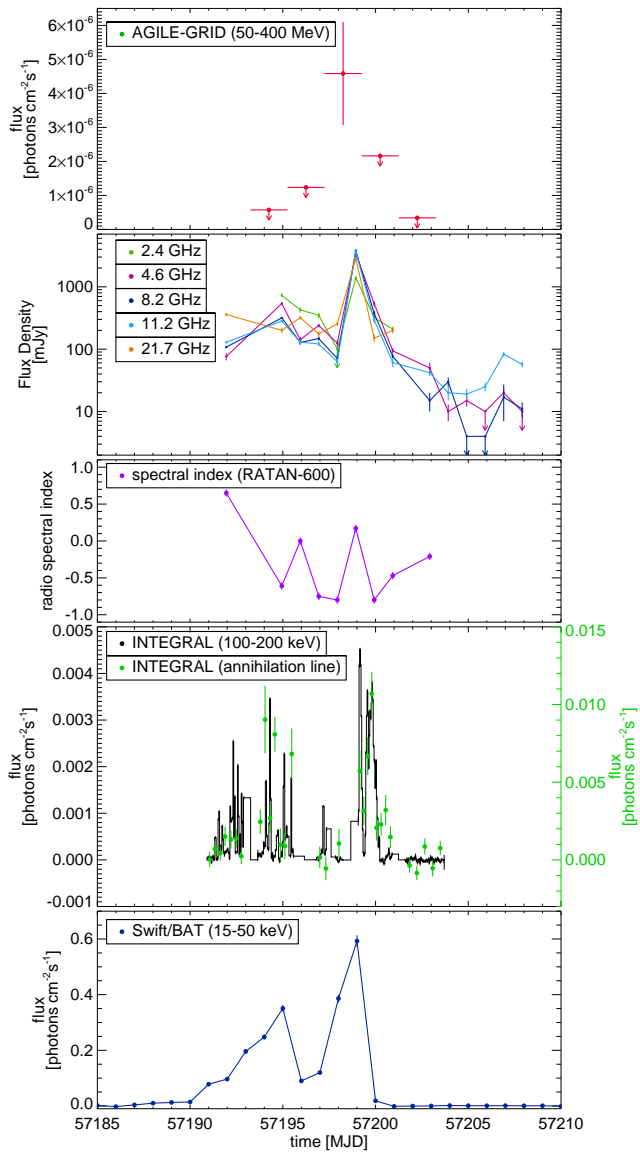}
	\vspace{-15pt}
	\caption{Multiwavelength light-curve throughout the June 2015 outburst of \vcyg. From top to bottom: \grid 50--400 MeV, 48 hours integration; RATAN-600 radio flux density  (2.4, 4.6, 8.2, 11.2 and 21.7 GHz); RATAN-600 radio spectral index (4-11 GHz); INTEGRAL/SPI continuum (100-200 keV, black histogram) and annihilation line (green points), data from \citet{siegert_2016}; \bat 15--50 keV, 1-day bin.}
	\label{fig:multiw_june2015}
  \end{center}
\end{figure}

\begin{figure*}
  \hspace*{-1.5cm}
    \begin{tabular}{cc}
    \includegraphics[width=\columnwidth]{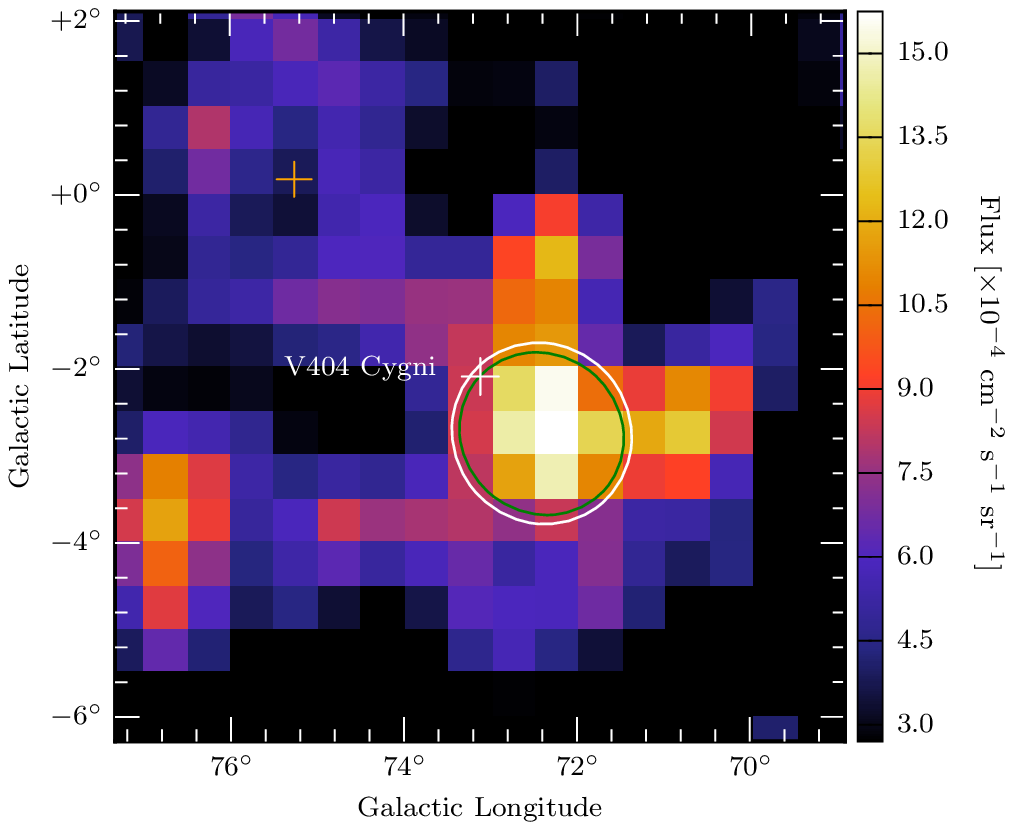}
    \includegraphics[width=\columnwidth]{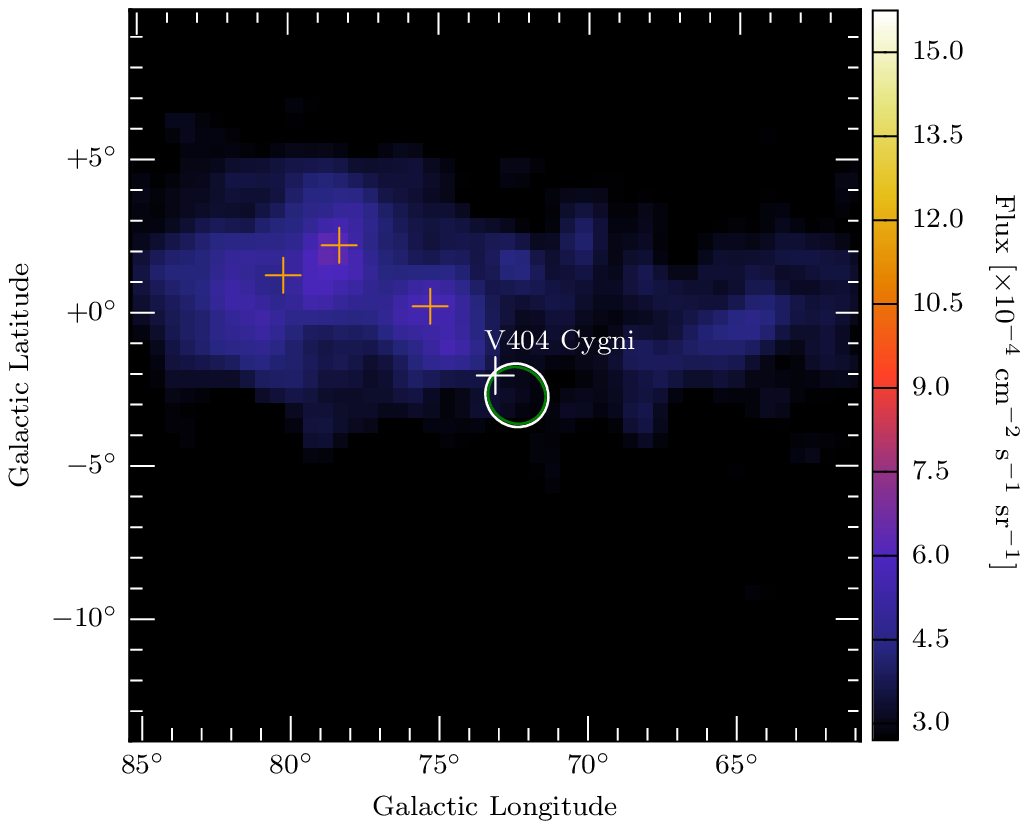}
     \end{tabular}
   	\caption{\textit{Left panel}: \grid \gray intensity map in Galactic coordinates with a three-pixel Gaussian smoothing. Photon energy: 50--400 MeV. Integration time: 2015-06-20 UT 06:00:00 -- 2015-06-30 UT 06:00:00. Pixel size: $0.5^{\circ}$. Green contour: $95\%$ confidence region. White contour: statistical + systematic ($0.1^{\circ}$) containment region. White cross: optical position of \vcyg. \textit{Right panel}: The quiescent phase of \vcyg -- with the same characteristics of the left panel, but different size -- from 2015-01-01 UT 12:00:00 to 2015-06-01 UT 12:00:00. The three pulsars included in the multi-source analysis are marked with magenta crosses (from left to right: PSR J2032+4127, PSR J2021+4026, and PSR J2021+3651). The white cross is the optical position of \vcyg. The white contour is the AGILE containment region of the flaring source (stat + syst). The color bar is the same for both the maps.}     
    \label{fig:agile_int_map}
\end{figure*}

In the 50--400 MeV energy band a detection was found in the time interval between 2015-06-24 UT 06:00:00 and 2015-06-26 UT 06:00:00 (MJD 57197.25--57199.25), at Galactic coordinates $(l, b) = (72.41^\circ, -2.75^\circ) \pm 0.97^\circ$ (stat.) $\pm~0.10^\circ$ (syst.), with TS = 18.1 (${\sim}4.3{\sigma}$) and a \gray flux $F_\gamma = (4.6 \pm 1.5) \times 10^{-6}$ \flx (see Fig.~\ref{fig:multiw_june2015} and \ref{fig:agile_fermi_june2015})\footnote{The systematic errors on flux measurements for AGILE have been quantified in 10\% of the total.}. In the \textit{left panel} of Fig.~\ref{fig:agile_int_map} the corresponding \grid \gray intensity map is shown, with the result of the best-fitting position. The nominal position of \vcyg is within the error ellipse of the AGILE \gray excess. The time correlation with the peak outburst phase observed in other wavelengths gives robustness to the association with \vcyg (see Section \ref{sec:multiwave}). For comparison, the \textit{right panel} shows the quiescent phase of \vcyg as detected between 2015-01-01 UT 12:00:00 and 2015-06-01 UT 12:00:00.

\begin{figure}[hb!]
  \begin{center}
	\includegraphics[width=\columnwidth]{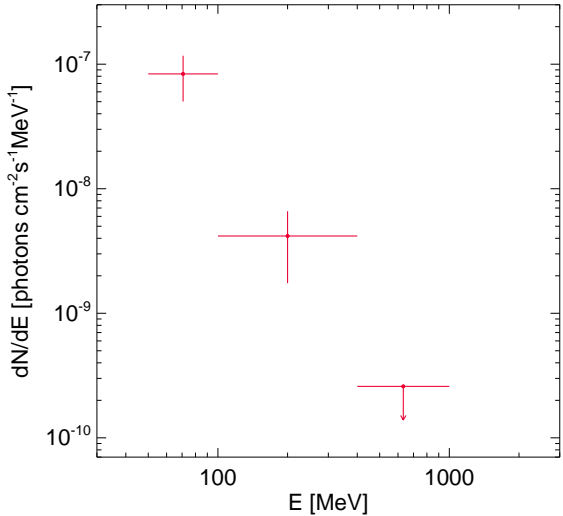}
	\caption{Differential \gray spectrum from \vcyg as detected by the \grid during the peak emission activity, 2015-06-24 UT 06:00:00 to 2015-06-26 UT 06:00:00 (MJD 57197.25--57199.25). The flux UL is a $68\%$ C. L. value.}
	\label{fig:agile_spectrum}
  \end{center}
\end{figure}

In Fig.~\ref{fig:agile_spectrum} the AGILE differential \gray spectrum (50 MeV -- 1 GeV) for \vcyg during the 48h peak \mod{activity is shown. No significant \gray emission is detected above 400 MeV.}

The \gray peak emission detected by AGILE is compatible in time with our analysis of  \lat data (TS =13.4, MJD 57198.75--59199.75) in the 60--400 MeV energy band. Furthermore, our findings are consistent with the \lat observations published in \citealp{loh_2016}. For a detailed description of our independent \lat data analysis see Appendix~\ref{appdx:fermi}. A comparison between the \grid and \lat observations and findings is presented in Appendix~\ref{appdx:agile-fermi}.

\begin{figure}[hb!]
  \begin{center}
	\includegraphics[width=\columnwidth]{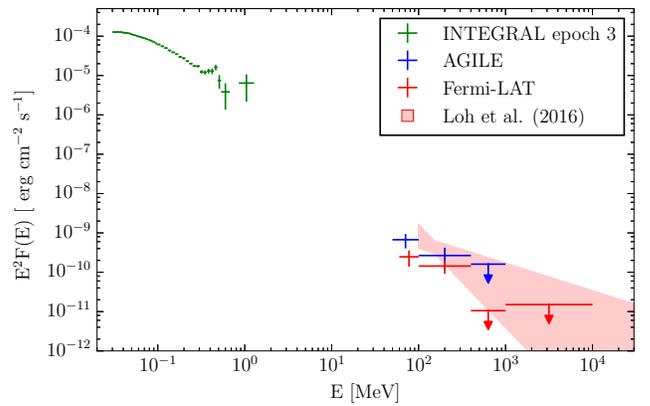}
	\caption{Multifrequency SED of \vcyg during the outburst phase. \textit{Green points}: INTEGRAL data, MJD $\sim$57199.1--57200.9 (epoch 3 in \citealp{siegert_2016}); \textit{Blue points}: AGILE data, MJD 57197.25--57199.25, \textit{Red points}: \lat data (see Appendix \ref{appdx:fermi}), MJD 57198.75--59199.75. \textit{Red shaded region}: 6h peak spectrum from \citet{loh_2016}, $\sim$MJD 57199.1 -- 57199.3. The flux ULs are $68\%$ C.L values.}
	\label{fig:nuFnu}
  \end{center}
\end{figure}

\begin{figure}[hb!]
  \begin{center}
	\includegraphics[width=\columnwidth]{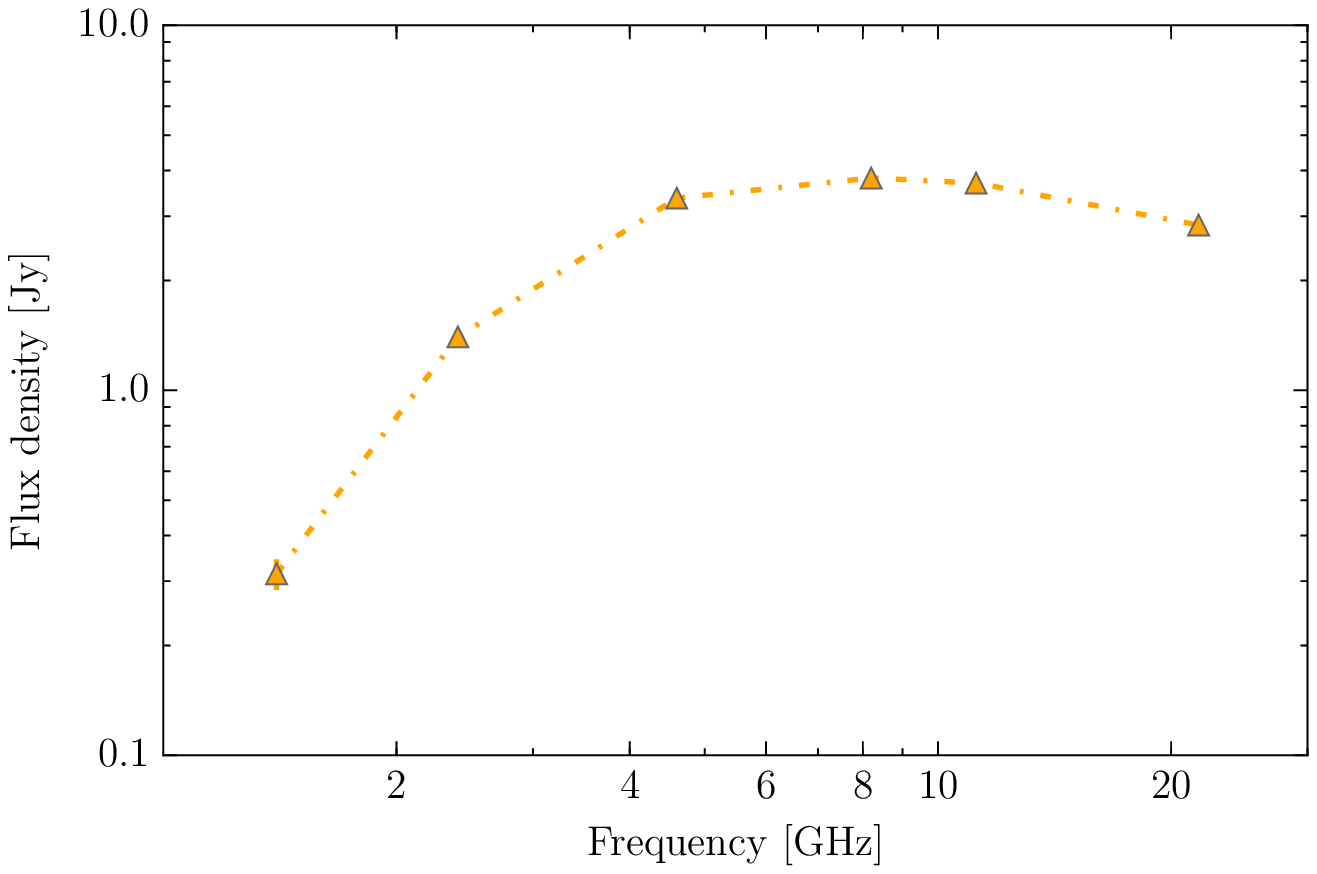}
	\caption{Radio spectrum of the outburst phase (MJD 57198.93) as observed by RATAN-600 \citep{trushkin_2015}.}
	\label{fig:radio_spec}
  \end{center}
\end{figure}

\section{Discussion}
\label{sec:discussion}
\subsection{AGILE results in a multi-wavelength context}
\label{sec:multiwave}
In Fig.~\ref{fig:multiw_june2015} we plotted the \grid light-curve together with published data from RATAN-600 \citep{trushkin_2015}, INTEGRAL/SPI \citep{siegert_2016}, and \bat\footnote{Public \bat light-curves: http://swift.gsfc.nasa.gov/results/transients/weak/V404Cyg/}. A plot showing the time evolution of the radio spectral index $\alpha$ (where $S \sim \nu^{\alpha}$ is the radio flux density), in the band 4-11 GHz, is shown, reporting a change between an optically thick ($\alpha > 0$) and an optically thin ($\alpha < 0$) regime. Such variations could indicate a multiple plasmoid ejection in the jet. In Fig.~\ref{fig:radio_spec} the radio spectrum, as detected by RATAN-600 during the outburst phase (MJD 57198.93), shows a change between an optically thick and an optically thin regime at ${\sim}4$ GHz.

There is a simultaneity of the peak emission detected by AGILE (and \lat), INTEGRAL/SPI (continuum 100--200 keV and annihilation emission), \bat (15--50 keV) and RATAN-600 (2.4--21.7 GHz). The \gray flare, detected by the \grid between 50 and 400 MeV, occurs at MJD  $57198.25 \pm 1$, with an integration time including the preparation and prompt phase of the prominent burst detected in radio and hard X-ray frequencies (2015 June 26, MJD 57199). The multiwavelength pattern is very similar to the Cyg X-3 evolution around the \gray flare \citep{tavani_2009b,fermi_2009,corbel_2012,piano_2012}, and it is consistent with a microquasar behavior, in which transient jets are responsible for the high-energy \gray emission (see Section \ref{sec:microquasar}). While the hard X-ray emission monitored by \bat in the 15--50 keV range is fully dominated by the disk-corona activity, the radio (RATAN-600) and soft/HE \gray radiation (INTEGRAL/SPI and \grid) clearly indicate the presence of a relativistic jet. INTEGRAL/SPI detected three time intervals of enhanced continuum emission at 100--200 keV (MJD: $\sim$57190.9--57192.9, $\sim$57193.6--57195.6, and $\sim$57199.1--57200.9). In particular, the second and third intervals show an evident correlation with the 511 keV pair-annihilation emission, suggesting the presence of a an unstable antiparticle outflow possibly related to the jet production \citep{siegert_2016}. Radio flux density evolution, as detected by RATAN-600, shows a first peak ($\sim$0.5 Jy at 4.6 GHz) coincident with the second burst activity observed by INTEGRAL/SPI ($\sim$57195.0). A giant radio flare ($\sim$3.4 Jy at 4.6 GHz) is detected during the last and brightest peak emission measured by INTEGRAL/SPI ($\sim$57199.2), which is consistent with the \gray flare detected by the \grid.

The AGILE observations are compatible with the \lat measurements, reported in \citealp{loh_2016} and in Appendix \ref{appdx:fermi}.
The contemporaneous burst observation of \vcyg by AGILE and \fermi gives statistical robustness to this episode, even though the detection significance is not impressive if considered individually. Moreover, the identification of \vcyg is secured by the time correlation with a strong outburst detected from the system at other wavelengths.

The multiwavelength behavior suggests that HE \gray activity is associated only with the highest activity phase of plasma ejection in the jet. In a jet scenario, the observed HE \grays must be produced outside the hot and dense corona region that is opaque (due to $e^+ ~ e^-$ pair production) for photon energies $E \gg m_e c^2$. According to this picture, HE \grays produced in the innermost part of the jet are converted to pair plasma. There is a continuous creation and annihilation of plasma close to the central source, forming a broad annihilation line \citep{siegert_2016, loh_2016}. Outside the coronal plasma, when the plasmoid moves away from the central source along the jet, HE \grays can propagate outwards without strong absorption (see, for a quantitative analysis applied to the specific case of Cyg X-3, \citealp{cerutti_2011}).

In Fig. \ref{fig:nuFnu} the Spectral Energy distribution (SED) at the outburst time is presented, showing INTEGRAL, AGILE and \lat data.

\subsection{Microquasar scenario}
\label{sec:microquasar}
As discussed above, the multifrequency emission pattern throughout the outburst clearly resembles the one observed for Cyg X-3 (\gray flares detected during X-ray spectral transitions and preceding giant radio outbursts, \citealp{corbel_2012,piano_2012}). The peak \gray isotropic luminosity for Cyg X-3 was found to be $L_{\gamma} \sim 10^{36}~\mathrm{erg~s^{-1}}$ (for $E_{\gamma} \geqslant 100$ MeV and a distance of 7--10 kpc, \citealp{tavani_2009b,fermi_2009}). It was found that both a leptonic (inverse Compton, IC) and hadronic ($\pi^0$-decay) scenario can account for the HE \gray emission, although IC processes can explain in a more natural way the \gray modulation and the multiwavelength links \citep{dubus_2010,piano_2012,sahakyan_2014}.

In \vcyg,  the simultaneous detection of HE \grays, $e^+ ~ e^-$ annihilation emission and a strong radio outburst can support a microquasar scenario with a dominant leptonic component responsible for the observed pattern of emission (at radio, hard-X/soft-\gray and HE \gray frequencies). 

In this work, for \vcyg we found a peak \gray luminosity $L_{\gamma} \sim 10^{35}~\mathrm{erg~s^{-1}}$, for $50 \leqslant E_{\gamma} \leqslant 400$ MeV and for a source distance of 2.39 kpc. According to the \gray observations by AGILE and \lat, $L_{\gamma} \sim 10^{-4} ~ L_{Edd}$ in \vcyg, which turns out to be a weaker \gray emitter with respect to Cyg X-3. Nevertheless, we have indications of a soft-spectrum HE \gray emission with no significant signal observed above 400 MeV. If we take into account the strong emission detected up to soft \grays by INTEGRAL and compare it with the low significance HE \gray flux detected by the \grid and \lat, we could ask ourselves whether the bulk \gray radiation is concentrated in the energy band between 1 and 50 MeV, a frequency window that current \gray detectors cannot observe. By a qualitative extrapolation to this energy range -- based on the observed trend (see Fig.~\ref{fig:nuFnu}) -- we could expect an energy flux around ${\sim}10^{-8}-10^{-6}$ \eflx. Is there a strong energy cut-off in the HE emission (indicating an upper limit to the Lorentz factor of the emitting particles in the jet)? Is the same spectral component responsible for the hard X-ray and \gray radiation? The next generation of \gray detectors, such as the e-ASTROGAM space mission \citep{tatischeff_2016}, will be able to explore this energy range with a good sensitivity, trying to disentangle the HE emission scenario for this kind of astrophysical sources.

\section{Acknowledgements}
\label{sec:acknowledgements} 
 
AGILE is an ASI space mission developed with programmatic support by INAF and INFN. This study was carried out with partial support through the ASI grant no. I/028/12/2.

The authors thank the anonymous referee for her/his stimulating comments on the manuscript and T. Siegert for providing the INTEGRAL data showed in Fig.~\ref{fig:multiw_june2015} and \ref{fig:nuFnu} (previously published in \citealp{siegert_2016}). 

\software{MSLA \citep{bulgarelli_2012}, Fermi Science Tools v10r0p52 (http://fermi.gsfc.nasa.gov), enrico (https://github.com/gammapy/enrico/)}

\appendix
 
  \section{\textit{Fermi}-LAT data analysis}
 \label{appdx:fermi}

In this appendix we analyzed the \lat \citep{atwood_2009} data using the \fermi Science Tools {\tt v10r0p5}\footnote{\tt http://fermi.gsfc.nasa.gov} and the user contributed package {\tt enrico}\footnote{\tt https://github.com/gammapy/enrico/}. We analyzed the data between 2015 June 17 UT 18:00:00 (MJD 57190.75) and 2015 July 02 UT 18:00:00 (MJD 57205.75), covering the peak activity period reported at other wavelengths.

We selected data with the \verb+P8R2_TRANSIENT_v16+ class in the widest energy range, given the expected faint short-lived activity of this binary in the gamma-ray domain. The data were centered at the position of \vcyg and extended within a region of interest (ROI) of $25^{\circ}$ radius. We adopted the last Galactic diffuse emission model (\verb+gll_iem_v06.fits+) and the isotropic model (\verb+iso_P8R2_SOURCE_V6_v06.txt+) in a likelihood analysis, and the 3rd point source catalog \verb+gll_psc_v16.fit+ \citep{acero_2016}. In the model, the Galactic diffuse and isotropic components are fixed to the values that we obtained in a long-time integration (1 month) preceding the active phase of June 2015. We selected PASS8 FRONT and BACK transient event class. We limited the reconstructed \verb+ZENITH_ANGLE+ to be less than $90^{\circ}$ (equivalent to an Earth albedo filtering angle of $75^{\circ}$) to strongly reduce \grays coming from the Earth's atmosphere. The good time intervals were selected so that the instrument rocking angle was lower than $52^{\circ}$. 

In the data modeling, we took into account nearby sources up to distances of $25^{\circ}$.
The analysis procedure was divided into two steps: in the first one, all sources closer than $6^{\circ}$ to \vcyg had all of their spectral parameters free, while sources further away had their parameters fixed. A likelihood analysis using the Minuit optimizer was run, determining the best fit parameters for our source and the nearby sources. In a second step, we fixed the spectral parameters for all the sources, except for \vcyg, to the ones found in the first step, and run the likelihood analysis again in order to obtain a refined fit. For \vcyg we used a simple power law model.

\begin{figure}
  \begin{center}
	\includegraphics[width=0.7\textwidth]{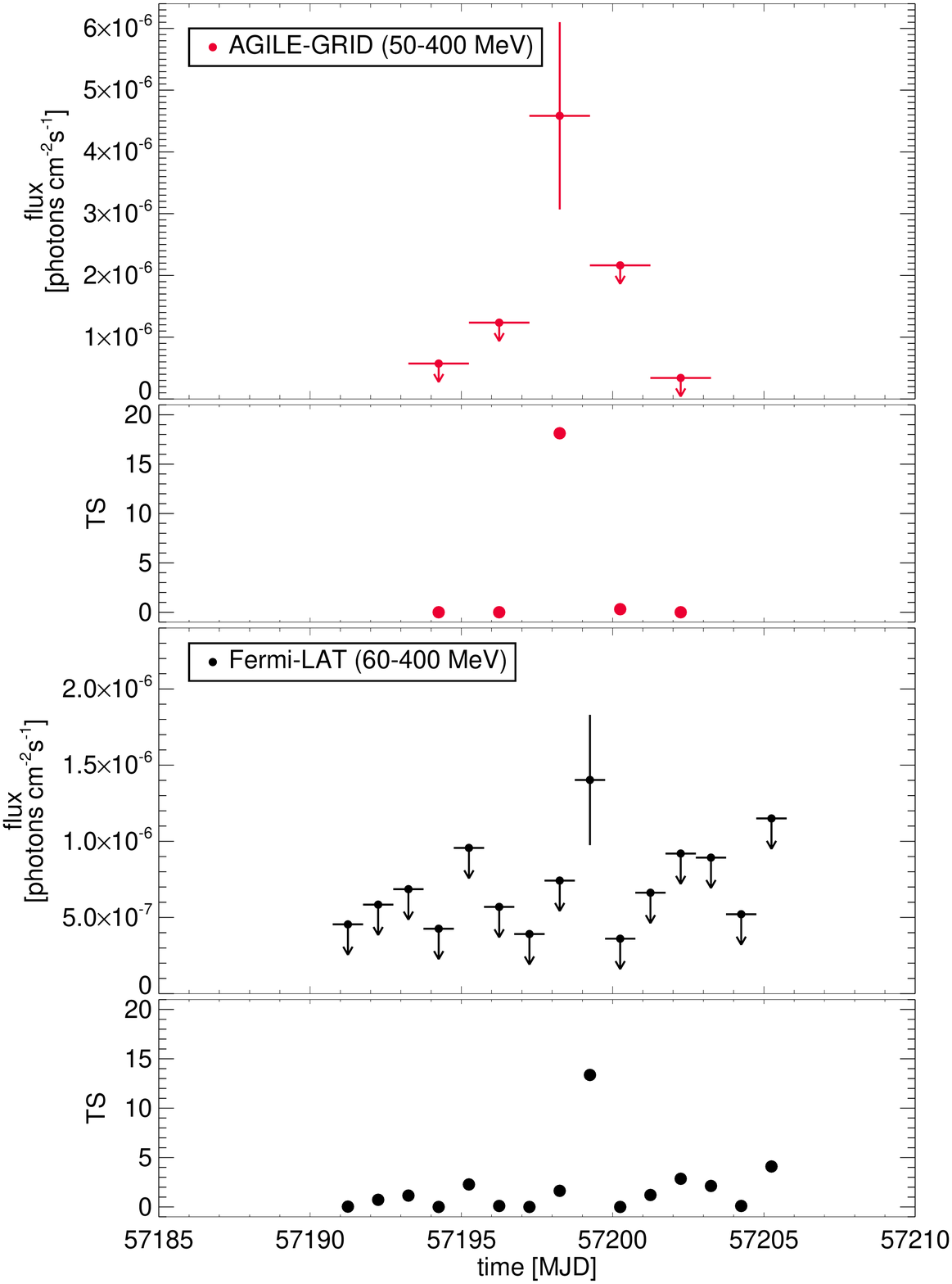}
	\caption{HE \gray light-curves across the \vcyg peak emission. From top to botton: \grid 48h-bin light-curve (50--400 MeV energy band); \grid TS of each time-bin; \lat 24h-bin light-curve (60--400 MeV energy band); \lat TS for each time-bin. For both AGILE and \fermi, flux error bars and flux ULs are $68\%$ C. L. values.}
	\label{fig:agile_fermi_june2015}
  \end{center}
\end{figure}

We produced light-curves with 24h time bins in two different energy bands: 60--400 MeV, and above 400 MeV. No significant emission was detected above 400 MeV. In the 60--400 MeV light-curve, instead, we obtained a hint of a detection at TS=13.4 ($\sim$3.7$\sigma$, see Fig.~\ref{fig:agile_fermi_june2015}) and a \gray flux $F_\gamma = (1.4 \pm 0.4) \times 10^{-6}$ \flx for the time integration between 2015-06-25 UT 18:00:00 and 2015-06-26 UT 18:00:00 (MJD 57198.75--59199.75). This result is fully consistent (in flux and integration time) with the 12h peak emission found by \citet{loh_2016} in their analysis of the \lat data in the 100 MeV -- 100 GeV energy range.

\subsection{AGILE--\textit{Fermi} comparison}
\label{appdx:agile-fermi} 

By comparing the \grid and our independent \lat analysis, we can note that both \gray instruments found a quasi-simultaneous peak emission with a time overlap of 12h. The \grid and \lat \gray fluxes observed at the peak emission are consistent with each other at 2$\sigma$ C.L..

AGILE and \fermi have different sky-scanning strategies and, consequently, they observe the target source at different times and for different duration in every single scan. Thus, if the \gray emission time is short (sub-hour variability as in the optical and X-ray frequencies), the detectors can miss part of the peak activity if the source is outside the field of view during the \gray activity.

\begin{figure}
  \begin{center}
	\includegraphics[width=0.7\textwidth]{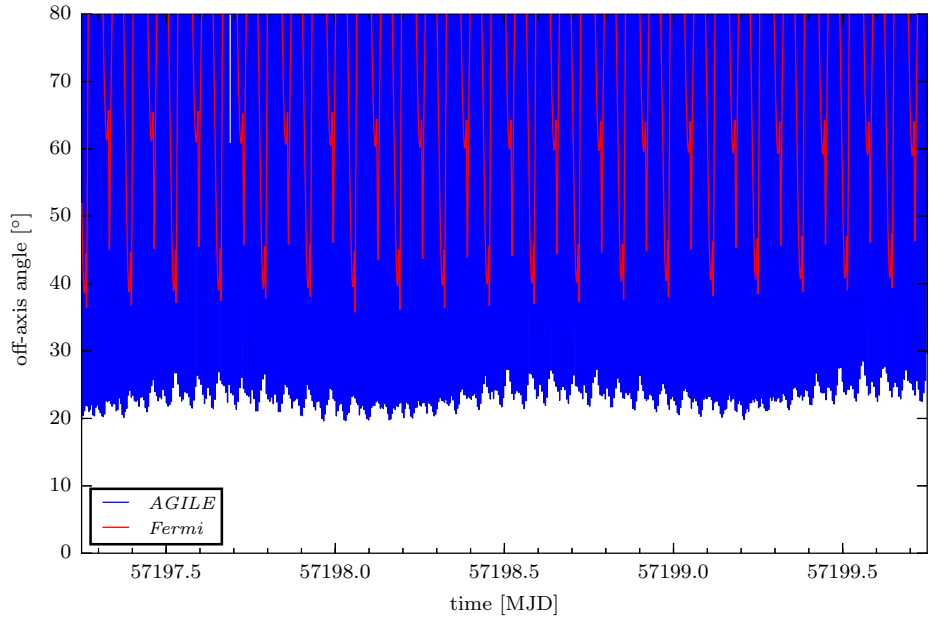}
	\caption{\grid (blue) and \lat (red) time evolution of the \vcyg off-axis angle, for the period 2015-06-24 UT 06:00:00 and 2015-06-26 UT 18:00:00 (MJD 57197.25--59199.75), union of the \gray peak emission time intervals found by AGILE and \fermi (see Fig.5~\ref{fig:agile_fermi_june2015}).}
	\label{fig:agile_fermi_exp}
  \end{center}
\end{figure}

Fig. \ref{fig:agile_fermi_exp} shows the \grid and \lat time evolution of the \vcyg off-axis angle during the period 2015-06-24 UT 06:00:00 and 2015-06-26 UT 18:00:00 (MJD 57197.25--59199.75), union of the \gray peak emission time intervals found by two instruments. As notices in other studies, AGILE and Fermi can have different exposure on specific target \citep{sabatini_2013,munar_2016}.

\end{document}